\documentclass[twocolumn,showpacs,preprintnumbers,amsmath,amssymb,prl,aps]{revtex4-2}
\usepackage{graphicx}
\usepackage{bm}
\usepackage{float} 
\usepackage{amsmath}
\usepackage{gensymb}
\usepackage{epstopdf}
\usepackage[usenames, dvipsnames]{color}

\begin{document}

\title{Behavior of gapped and ungapped Dirac cones in an antiferromagnetic topological metal, SmBi}

\author{Anup Pradhan Sakhya$^1$}
\author{Shiv Kumar$^2$}
\author{Arindam Pramanik$^1$}
\author{Ram Prakash Pandeya$^1$}
\author{Rahul Verma$^1$}
\author{Bahadur Singh$^1$}
\author{Sawani Datta$^1$}
\author{Souvik Sasmal$^1$}
\author{Rajib Mondal$^1$}
\author{Eike F. Schwier$^2$}
\author{Kenya Shimada$^2$}
\author{A. Thamizhavel$^1$}
\author{Kalobaran Maiti$^1$}
\altaffiliation{Corresponding author: kbmaiti@tifr.res.in}

\affiliation{$^1$Department of Condensed Matter Physics and Materials Science, Tata Institute of Fundamental Research, Homi Bhabha Road, Colaba, Mumbai - 400005, INDIA}
\affiliation{$^2$Hiroshima Synchrotron Radiation Center, Hiroshima University, 2-313 Kagamiyama, Higashi-Hiroshima 739-0046, Japan.}

\begin{abstract}
We studied the behavior of nontrivial Dirac fermion states in an antiferromagnetic metal SmBi using angle-resolved photoemission spectroscopy (ARPES). The experimental results exhibit multiple Fermi pockets around $\overline{\Gamma}$ and $\overline{M}$ points along with a band inversion in the spectrum along the $\overline{\Gamma}$-$\overline{M}$ line consistent with the density functional theory results. In addition, ARPES data reveal Dirac cones at $\overline{\Gamma}$ and $\overline{M}$ points within the energy gap of the bulk bands. The Dirac cone at $\overline{M}$ exhibit a distinct Dirac point and is intense in the high photon energy data while the Dirac cone at $\overline{\Gamma}$ is intense at low photon energies. Employing ultra-high-resolution ARPES, we discover destruction of a Fermi surface constituted by the surface states across the Ne\'{e}l temperature of 9 K. Interestingly, the Dirac cone at $\overline{\Gamma}$ is found to be gapped at 15 K and the behavior remains similar across the magnetic transition. These results reveal complex momentum dependent gap formation and fermi surface destruction across magnetic transition in an exotic correlated topological material; the interplay between magnetism and topology in this system calls for ideas beyond existing theoretical models.
\end{abstract}

\date{\today}

\maketitle

\section{Introduction}

Topological insulators (TIs) attracted much attention due to their potential for quantum science and device applications \cite{Hasan}. In these materials, spin-orbit coupling (SOC) leads to band inversion and the topological surface states (TSS) possessing Dirac-cone-like dispersion and vortex-like spin texture appear inside the inverted bandgap in the presence of time-reversal and/or mirror symmetry \cite{Kane}. In topological (Dirac and Weyl) semimetals, the valence and conduction bands touch at discrete points or extended lines within the bulk Brillouin zone (BBZ). In Dirac semimetals, the doubly degenerate bands respecting inversion and time-reversal symmetries cross at the Dirac point (DP). Breaking of time-reversal or inversion symmetry lifts the band degeneracy to realize Weyl semimetals \cite{suyang, Vish}. Research in this field significantly involves the discovery of new materials where such physics can be realized. Recently, RSb/RBi (R = rare earth) family of materials are studied extensively reporting Dirac states and giant magnetoresistance \cite{zeng, vaithee, Yang, Niu, Jayita, Kaminski, Yuan, Lou, Feng, Hosen}. E.g., LaBi exhibits three Dirac cones; one at $\overline{\Gamma}$ and two at $\overline{M}$ of the surface Brillouin zone (SBZ) \cite{Jayita}. The Dirac cone at $\overline{\Gamma}$ is gapped and exhibit asymmetric mass acquisition \cite{Kaminski}. Another study \cite{Lou} reported one Dirac cone at $\overline{M}$.

Electron correlation is expected to influence the effective mass of Dirac fermions leading to exotic properties. A Kondo insulator, SmB$_6$ was proposed to be in this class and host TSS within the Kondo gap \cite{Coleman}. Extensive studies reported interesting surface/bulk electronic structure of SmB$_6$ although its topological behavior is widely debated \cite{Neupane,Jiang,Xu,Denlinger,anupsmb6}. The properties of (Pr, Sm)Sb and (Pr, Sm)Bi were studied extensively to probe the origin of extreme magneto-resistance in these materials \cite{Yang1,yogesh}. SmBi belongs to this class and predicted to host TSS from band structure calculations\cite{Duan} while ARPES study \cite{Yang} does not show evidence of such TSS. A recent study of the magnetic and transport properties of SmBi show an antiferromagnetic (AFM) transition at 9K, Kondo-type resistivity and signature of non-trivial Berry phase \cite{Anup}. Clearly, SmBi can be a candidate material for the study of interplay between topology and magnetism. Here, we studied the electronic structure of SmBi employing high resolution ARPES and first-principles calculations, and discover coexistence of widely varying Dirac fermions and their unusual evolution across magnetic transition.

\section{Experimental and Computational Details}

High quality single crystals of SmBi were grown by flux method \cite{Anup}. ARPES measurements were performed on (001) surface prepared by cleaving at 6K using ultrahigh-resolution $\mu$-laser ARPES setup ($h\nu$ = 6.3 eV; R4000 analyzer) and $s$ polarization geometry at Hiroshima Synchrotron Radiation Center (HiSOR) [energy resolution $<$260$\mu$eV, angular resolution $<$0.05$\degree$] \cite{ShimadaLaser}. The synchrotron-ARPES measurements were done at BL-1 beamline, HiSOR \cite{ShimadaSynch} [energy-resolutions = 20meV, angle-resolution = 0.1$^\circ$]. The Fermi level, $\epsilon_F$ was determined using a gold film evaporated onto the sample holder. Electronic structure calculations were performed using the projector augmented wave method as implemented in the Vienna ab-initio simulation package (VASP)~\cite{dft,vasp}. The generalized gradient approximation (GGA) was employed to include the exchange-correlation effects~\cite{gga}. For magnetic solutions, on-site Coulomb interaction was added for Sm 4$f$ electrons within the GGA+U scheme~\cite{ggaU}. An energy cutoff of 480 eV was used for the plane-wave basis set and a 11$\times$11$\times$11 $k$-mesh was considered for BBZ integration. Topological properties were obtained by employing the materials specific tight-binding Hamiltonian using the WannierTools package~\cite{wannier,wt}.

\section{Results and Discussions}

\begin{figure}
\includegraphics[width=0.5\textwidth]{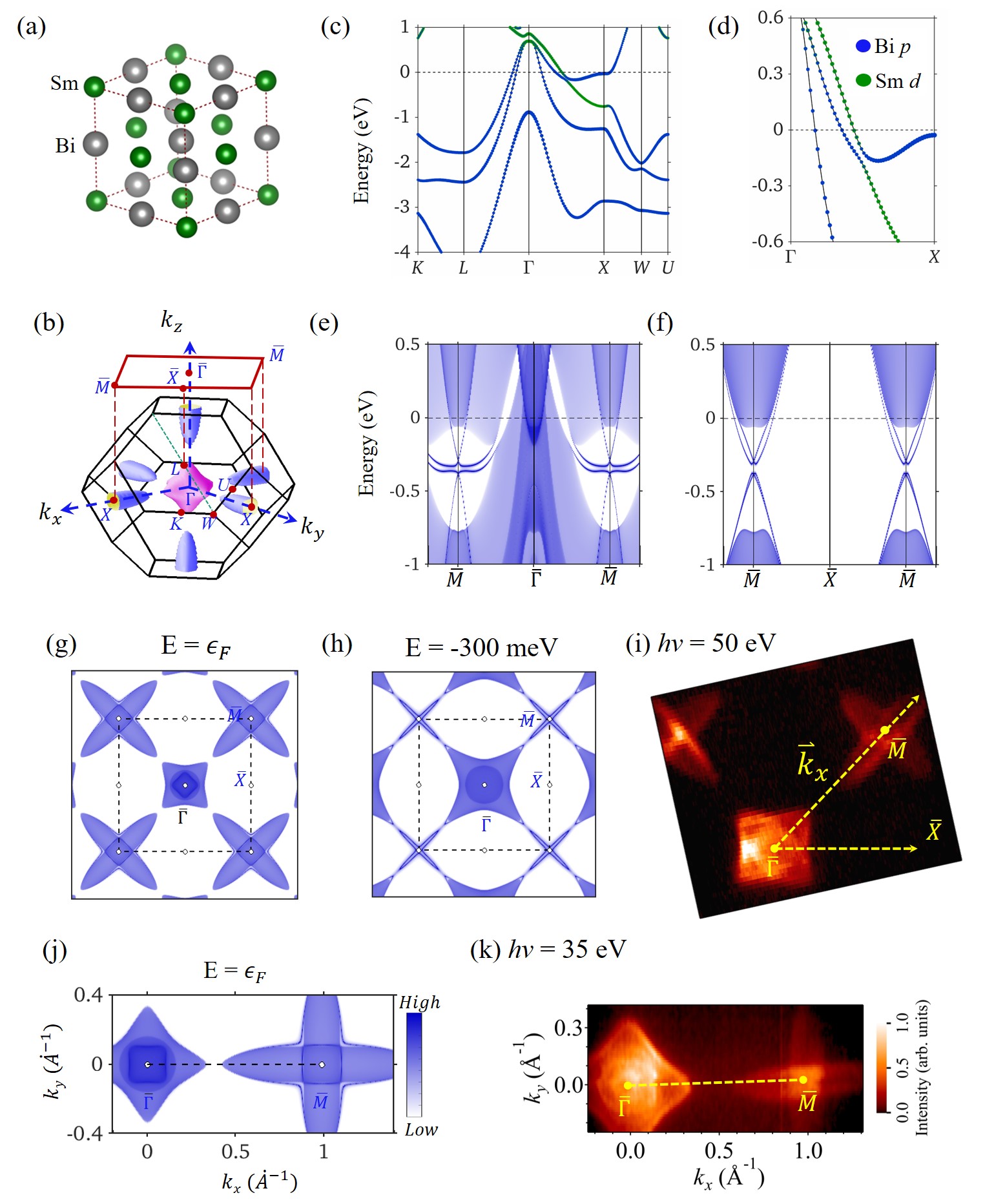}
\caption{(a) Crystal structure of SmBi. (b) Fermi surfaces within BBZ and (001) surface projected SBZ. There is a spherical hole pocket (not seen) inside the star-shaped hole pockets at $\Gamma$. (c) Band structure of non-magnetic SmBi including SOC. The blue and green markers highlight B 6$p$ and Sm 5$d$ states. (d) Closeup of bands along $\Gamma-X$ line exhibiting bulk band inversion. Calculated (001) surface band structure along  (e) $\overline{M}$-$\overline{\Gamma}$-$\overline{M}$ and (f)  $\overline{M}$-$\overline{X}$-$\overline{M}$ of the SBZ. Two Dirac cones at $\overline{M}$ point are resolved. Dirac states at $\overline{\Gamma}$ point are masked with bulk band projections. Calculated surface band contours at (g) Fermi energy ($E = \epsilon_f$) and (h) $E$ = -300 meV (near Dirac point). (i) Experimental Fermi surface collected with a photon energy of  50 eV. (j) Calculated and (k) measured surface Fermi contours along $\overline{\Gamma}$-$\overline{M}$ direction.
}
\label{figure1}
\end{figure}

The crystal structure of SmBi (space group \emph{Fm$\overline3$m}) and the calculated bulk Fermi surfaces (FS) are shown in Figs. \ref{figure1}(a) and (b), respectively. In Fig. \ref{figure1}(b), the high symmetry points of BBZ and their projection on (001) surface are also shown. The calculated bulk band structure in Fig. \ref{figure1}(c) exhibit three bands crossing $\epsilon_F$ along $\Gamma-X$ line; the closeup of these bands [see Fig. \ref{figure1}(d)] illustrates a band inversion between Bi $p$ and Sm $d$ derived bands at $X$ points. Antiband crossing features are resolved along $\Gamma-X$ line indicating that valence and conduction bands are locally separated at each $k$ point. The $Z_2$ topological invariant obtained by calculating inversion eigenvalues of the occupied bands at eight time-reversal invariant momenta is 1 suggesting presence of non-trivial topological phase in this system \cite{Kane, Duan}. Figs.~\ref{figure1}(e)-(f) show the calculated (001) surface band structure along $\overline{M}$-$\overline{\Gamma}$-$\overline{M}$ and $\overline{M}$-$\overline{X}$-$\overline{M}$ lines. Two Dirac cones are resolved at $\overline{M}$ consistent with the projection of two $X$ points with inverted bulk bands. An additional Dirac cone is expected at $\overline{\Gamma}$ although it is masked with the bulk projections. The calculated surface Fermi contours are shown in Fig. \ref{figure1}(g). It exhibits a star-shaped Fermi pocket near $\overline{M}$ and a circular hole pocket around $\overline{\Gamma}$. The square-shaped pocketaround $\overline{\Gamma}$ is the projection of electron pockets at $X$ of BBZ. Ellipsoidal pockets around $\overline{M}$ are the projection of the electron pockets of $X$-points. The surface band contours close to DP ($E$ = -300 meV) in Fig.~\ref{figure1}(h) unfold the surface bands.

Experimental FSs measured using 50 eV and 35 eV photon energies are shown in Figs. \ref{figure1}(i) and (k). We observe one circular-shaped FS inside a star-shaped FS at $\overline{\Gamma}$ representing the two-hole pockets consistent with the calculations. Overlapping elliptical Fermi pockets at $\overline{M}$ are the projection of electron pockets. There is an additional intense circle inside the bulk circular hole pocket at $\overline{\Gamma}$ in the 35 eV data. The calculated and measured Fermi contours are in substantial agreement as shown in Figs.~ \ref{figure1}(j) and (k).

\begin{figure}
 \includegraphics[width=1.0\linewidth]{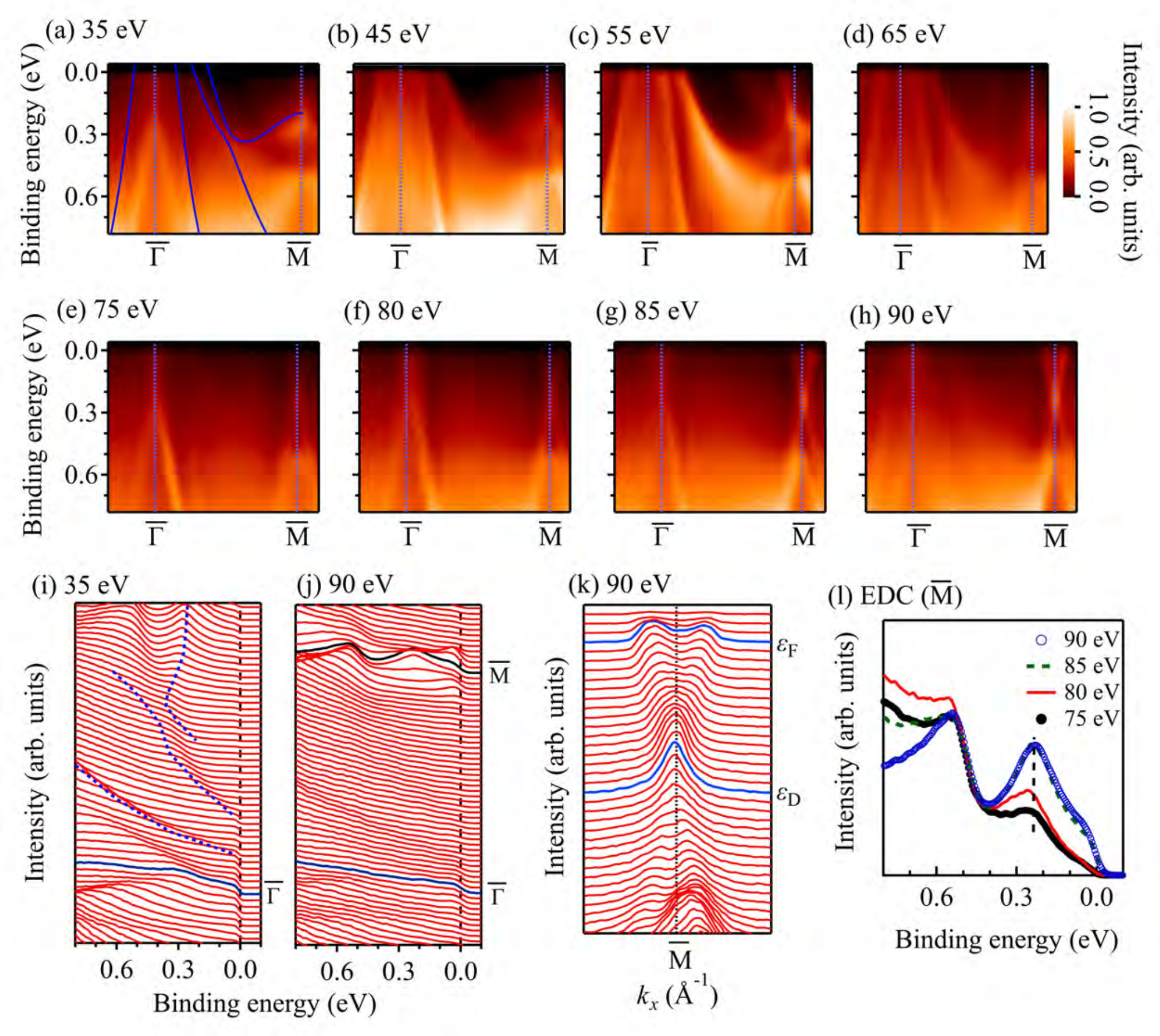}
\caption{(a)-(h) ARPES data along $\overline{\Gamma}$-$\overline{M}$ line measured with photon energies 35-90 eV. Solid line in (a) are the DFT results. (i) EDCs of the 35 eV ARPES data near $\overline{\Gamma}$ showing the band inversion and a distinct Dirac cone at $\overline{\Gamma}$. (j) EDCs and (k) MDCs of 90 eV ARPES data show Dirac cone at $\overline{M}$. (l) EDC at $\overline{M}$ at different photon energies exhibiting Dirac node (vertical dashed line) almost at the same energy.}
\label{Figure2}
\end{figure}

In Fig. \ref{Figure2}(a)-(h), we show the energy bands measured along $\overline{\Gamma}-\overline{M}$ line using varied photon energies. The calculated bulk band structure are shown by lines in Fig. \ref{Figure2}(a) exhibiting good description. The bands corresponding to two-hole pockets around $\overline{\Gamma}$ and an electron pocket at  $\overline{M}$ are evident in the figure. Outer hole band and electron band in Fig. \ref{Figure2}(a) and Fig. \ref{Figure2}(i) exhibit band-inversion along $\overline{\Gamma}-\overline{M}$ line due to Bi 6$p$-Sm 5$d$ overlap consistent with the DFT results.

Interestingly, 35eV ARPES data show a Dirac cone around $\overline{\Gamma}$. Another Dirac cone at $\overline{M}$ is observed in high photon energies. The dependence of intensity of Dirac bands to the photon energy arises from the photoemission matrix element effects \cite{yeh,Borisenko}. Clearly, the properties of the Dirac cones at  $\overline{\Gamma}$ and  $\overline{M}$ are significantly different. Earlier studies of same material class reported two Dirac cones at $\overline{M}$ in some cases\cite{Yang,Niu,Jayita} while one Dirac cone in others \cite{Lou,Feng}. We identify one cone in the 90 eV data shown in Fig. \ref{Figure2}(j) and the corresponding momentum distribution curves (MDCs) in Fig. \ref{Figure2}(k), which is similar to the latter case and/or the Dirac points in SmBi are too close to be detected distinctly. The DP (vertical dashed line) in the EDCs at $\overline{M}$ shown in Fig. \ref{Figure2}(l) does not shift with the change in photon energy ($\Gamma$ on $k_z$-axis appears at 51 eV and 85 eV assuming inner potential, 14 eV). Absence of $k_z$-dependence of DP suggests two-dimensional behavior of Dirac fermions.

\begin{figure}
 \includegraphics[width=1.0\linewidth]{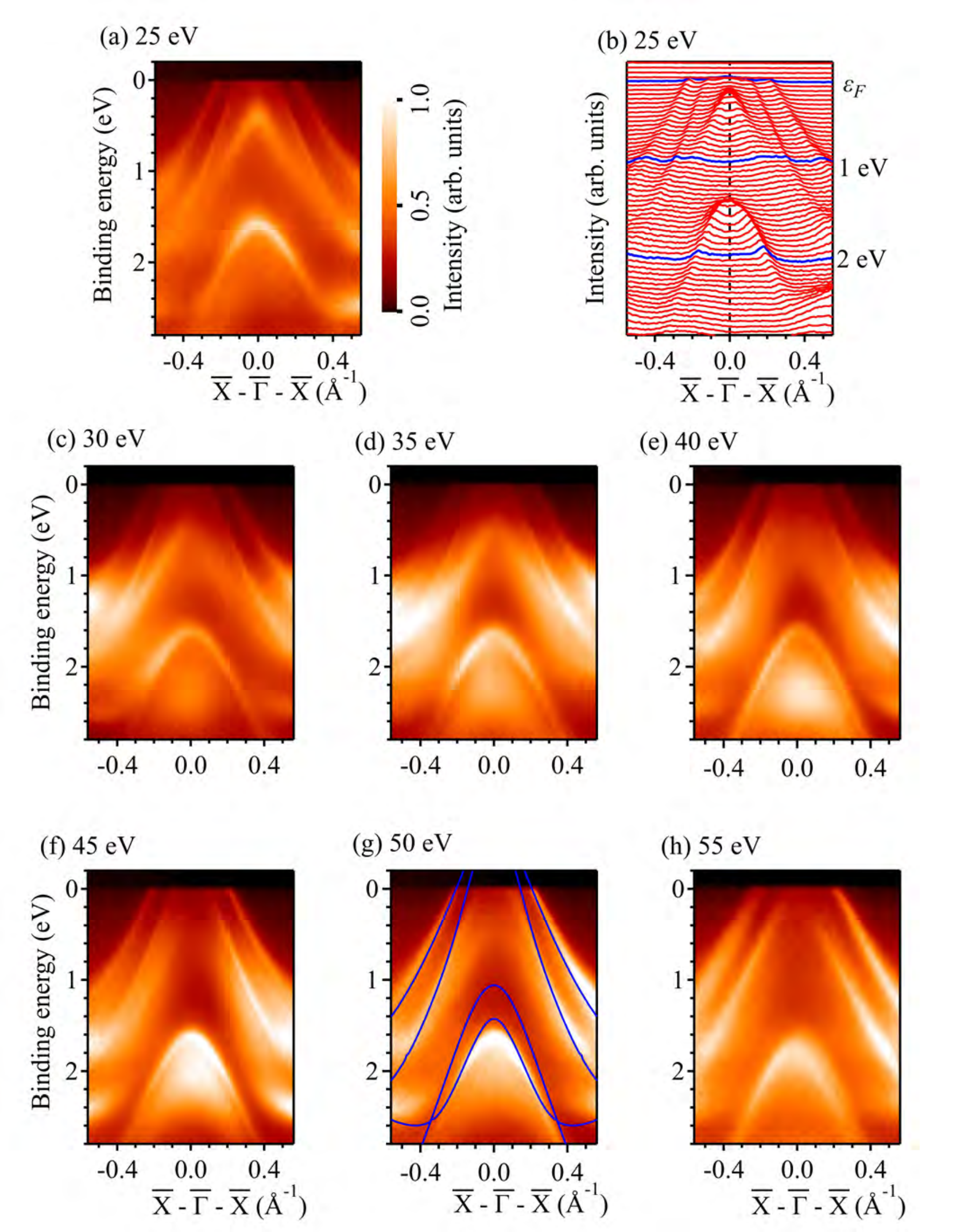}
\caption{(a) ARPES spectra measured at 30 K along the $\overline{\Gamma}$-$\overline{X}$ vector using 25 eV photon energy and (b) it's corresponding MDCs. (c-h) ARPES spectra using similar conditions and photon energies varying from 30 - 55 eV. Solid lines in (g) are the DFT results.}
\label{figure3}
\end{figure}

The band structure along $\overline{\Gamma}-\overline{X}$ line measured using 25 eV photon energy is shown in Fig. \ref{figure3}(a) and corresponding MDCs are shown in Fig. \ref{figure3}(b). The ARPES data at 30 - 55 eV are shown in Figs. \ref{figure3}(c)-(h). Similar to Fig. \ref{Figure2}, we observed two distinct bands cross $\epsilon_F$ forming two hole pockets around $\overline{\Gamma}$. Intense bands beyond 1.5 eV are also observed. These results represent the bulk band structure and are in good agreement with the DFT results as shown by solid lines superimposed on the experimental data. The bands forming a Dirac-cone at $\overline{\Gamma}$ in addition to the bulk bands are clearly visible in the 25 - 35 eV photon energy spectra. While the intensity of the upper cone part is weak, the lower part is intense at 25 eV and gradually becomes weaker with the increase in photon energy due to the matrix element effect \cite{yeh}.

\begin{figure}
 \includegraphics[width=1.0\linewidth]{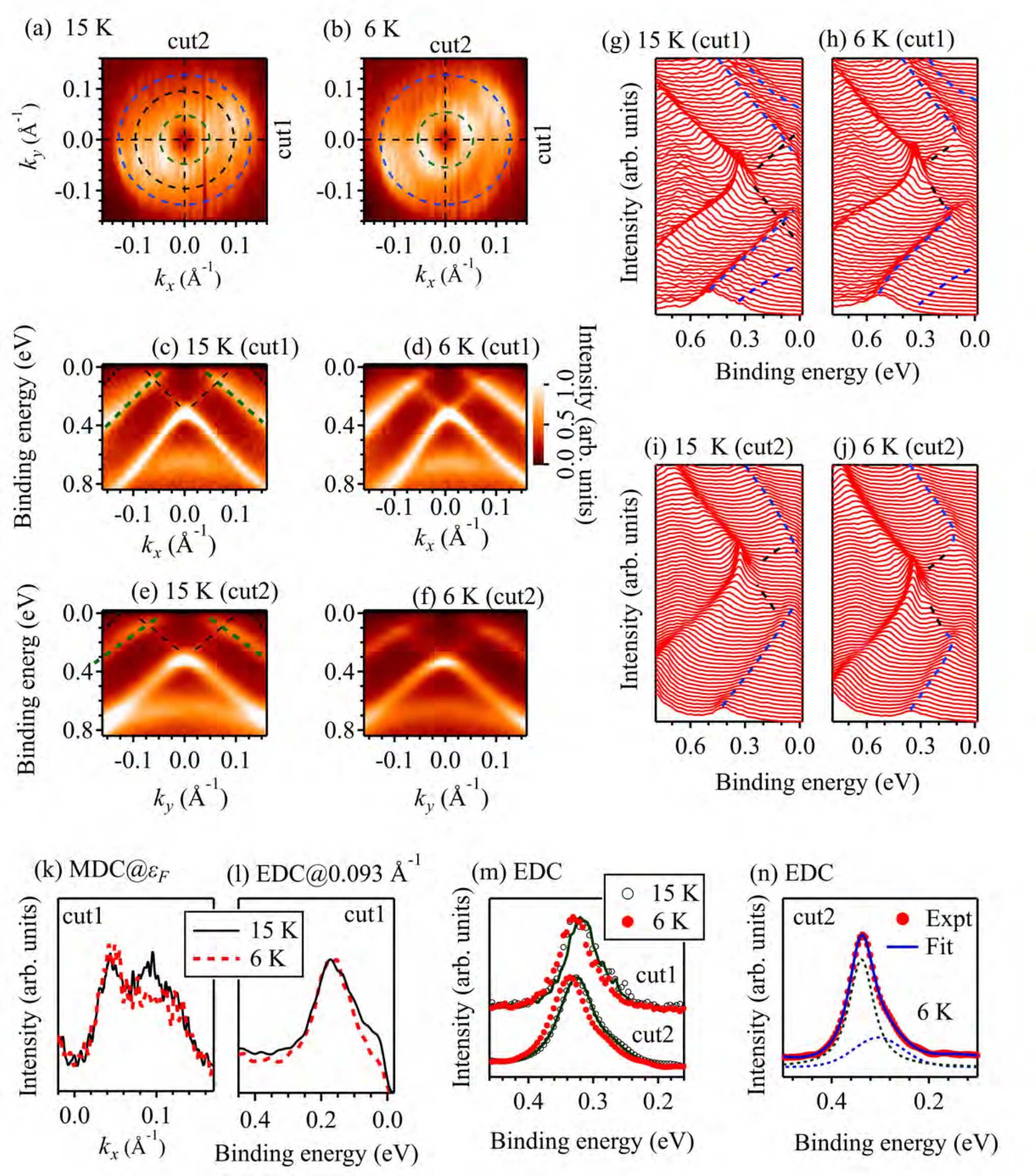}
\caption{Fermi surface at (a) 15K and (b) 6K in the ultrahigh resolution laser ARPES data. Band dispersions along cut1 at (c) 15K and (d) 6K, and along cut2 at (e) 15K and (f) 6K. Corresponding EDCs along cut1 at (g) 15K and (h) 6K, and along cut2 at (i) 15K and (j) 6K; dashed lines are the hand drawn curves. (k) MDCs at $\epsilon_F$ at 15K (solid line) and 6K (dashed line) for cut1. (l) EDCs at $k_F$ (= 0.093 \AA$^{-1}$) at 15K (solid line) and 6K (dashed line). (m) EDCs at $\overline{\Gamma}$ at 15K  (open circles) and 6K (solid circles). Lines overlaid on the 15K data represent the 6K data shifted by 11 meV. (n) Fit of the EDC at $\overline{\Gamma}$ along cut2 at 6K showing an energy gap at the Dirac point. Dotted lines are the component peaks.}
\label{figure4}
\end{figure}

In Fig. \ref{figure4}, we show the ultrahigh-resolution laser ARPES data. The data at 15K exhibit three circular Fermi pockets as marked by the dashed lines. The outermost circle represents the bulk hole pocket as seen in the DFT results and the synchrotron data in Fig. \ref{figure1}. Signature of surface Fermi pocket is also observed in the synchrotron ARPES data [see Fig. \ref{figure1}(k)]; ultrahigh resolution laser ARPES data resolved distinct signature of two pockets due to the surface bands. Interestingly, the outer surface pocket is absent in the 6K data suggesting destruction of the surface states derived Fermi pockets across the magnetic transition.

The bands along cut1 and cut2 are shown in Figs. \ref{figure4}(c)-(f). At 15K, intense bands cross $\epsilon_F$ that formed the innermost Fermi pocket. A Dirac cone is seen with DP around 320 meV at $\overline{\Gamma}$. The lower part of the cone is much intense compared to the upper part. Despite weaker intensity, distinct linear bands are observed to cross $\epsilon_F$ and form the middle Fermi pocket as evident in Figs. \ref{figure4}(c) and (e). This is verified by the MDCs at $\epsilon_F$ in Fig. \ref{figure4}(k); three distinct peaks in the 15K data reflects the signature of three Fermi pockets. The middle peak around 0.093 \AA$^{-1}$ is absent in the 6K data as also observed in the FS plot in Fig. \ref{figure4}(b). The EDCs in Figs. \ref{figure4}(h) and (j) suggest that the linear Dirac bands form an energy gap below the Ne\'{e}l temperature. The EDCs at 15K shown in Figs. \ref{figure4}(g) and (i) do not exhibit this property. To confirm the gap formation, we plot the EDCs at 15K and 6K at $k_F$ (= 0.093 \AA$^{-1}$) in Fig. \ref{figure4}(l) which show significant intensity loss at $\epsilon_F$ at 6K compared to the 15K data. These results demonstrate the destruction of the Fermi pocket across the antiferromagnetic transition; a signature of complex scenario of magnetism of the topological states.

We now analyse the spectral intensities at DP in Figs. \ref{figure4}(m). The EDCs show distinct signature of two peak structure as seen in the fit using Voigt functions for the 6K data along cut2 in Fig. \ref{figure4}(n). The energy gap is estimated to be about 25 ($\pm$5) meV at DP. The results are quite similar for the data along cut1 too. Interestingly, if we shift the 6K data by 11 meV towards lower binding energy (see solid line in Fig. \ref{figure4}(m)), it superimpose on the 15K data well. This suggests that the properties of the Dirac fermions near DP are not significantly influenced by the antiferromagnetic order.

\begin{figure}
 \includegraphics[width=1.0\linewidth]{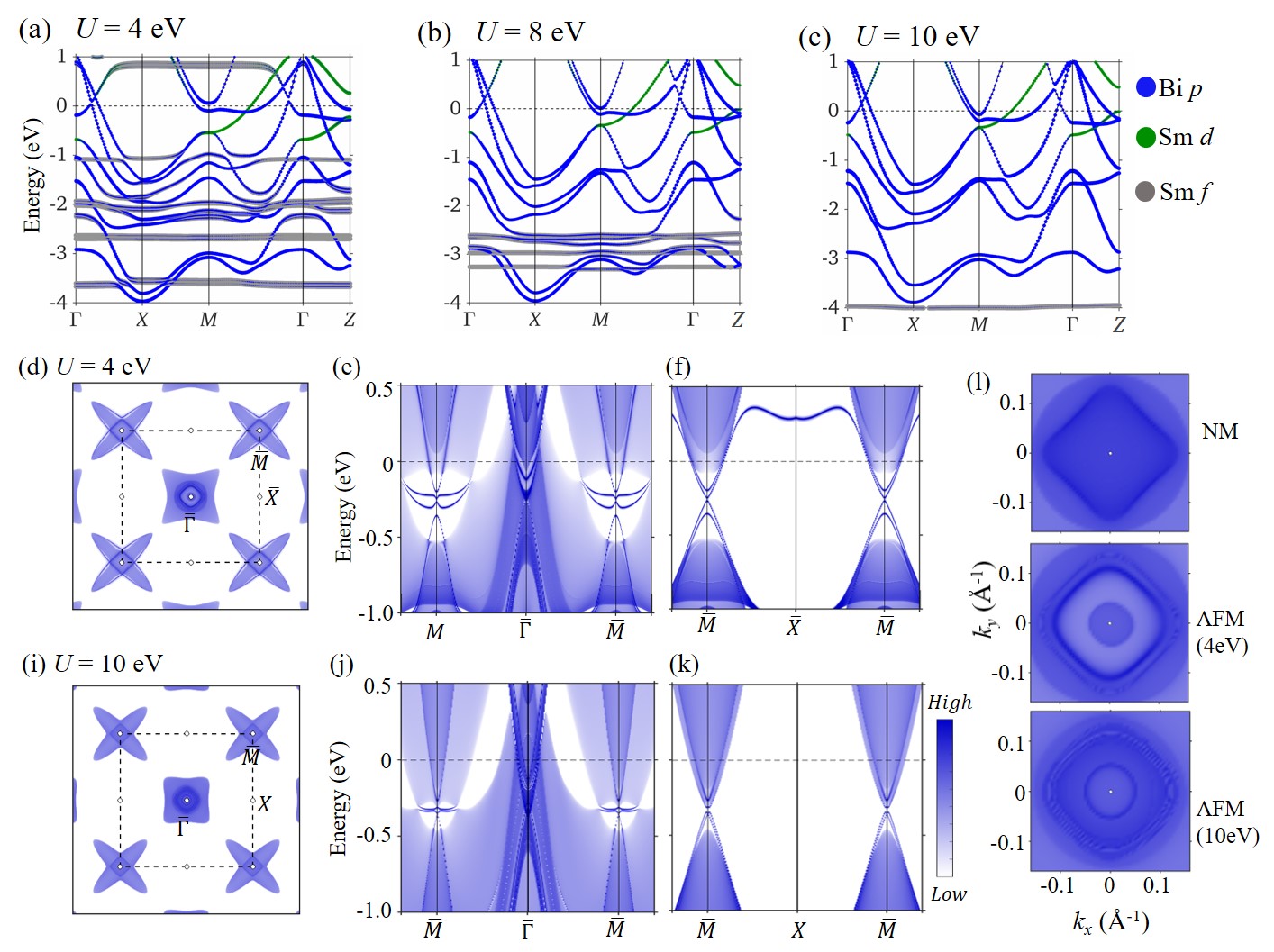}
\caption{Band structure of AFM SmBi in the conventional bulk Brillouin zone for $U_{eff}$ = (a) 4 eV, (b) 8 eV and (c) 10 eV. The Fermi level is adjusted to match the experimentally observed Dirac point at $\overline{\Gamma}$. (d) Calculated (001) surface projected Fermi contours. $E-k$ dispersions along (e) $\overline{M}-\overline{\Gamma}-\overline{M}$ and (f) $\overline{M}-\overline{X}-\overline{M}$ line for $U_{eff}$ = 4 eV. (i) - (k) are the same for $U_{eff}$ = 10 eV. (l) Fermi surface around $\overline{\Gamma}$ for non-magnetic and AFM solutions ($U_{eff}$ = 4 eV, 10 eV).}
\label{figure5}
\end{figure}

To investigate the AFM state theoretically, we show calculated energy bands in Fig. \ref{figure5}. We find that the inverted energy gap at $\overline{M}$ gradually reduces with enhancement in effective electron correlation strength, $U_{eff}$ among Sm 4$f$ electrons. However, the overall properties of the bulk bands in Figs. \ref{figure5}(a) - (c) are similar to the nonmagnetic case. The FS contours for $U_{eff}$ = 4eV and 10eV are shown in Fig. \ref{figure5}(d) and (i). In Figs. \ref{figure5}(e) and (f), and (j) and (k), the Dirac states at $\overline{M}$ form an energy dispersion similar to the non-magnetic phase along with a gap at DP due to time-reversal symmetry breaking at the surface. Interestingly, a large gap appears at DP at $\overline{\Gamma}$ shown in Fig. \ref{figure5}(e) and (j), which was absent in non-magnetic phase. To compare FS with the ultrahigh resolution data in Fig. \ref{figure4}, we show FS around $\overline{\Gamma}$ in Fig. \ref{figure5}(l); the square-shaped FS is the projection of electron pockets at $X$-point. There are three Fermi pockets around $\overline{\Gamma}$ in the non-magnetic solution; the surface contributions are not clearly visible due to heavy bulk band projections. The AFM solution show two circular pockets as observed in the experiment indicating destruction of one surface FS across the AFM order. These results suggest that the DP gap at $\overline{\Gamma}$ and the destruction of FS observed in experiments may be arising from antiferromagnetic interactions present in the system.

\section{Conclusions}

In summary, we studied the electronic structure of a topological antiferromagnetic metal, SmBi using high-resolution ARPES and density functional theoretical calculations. The energy bands along $\overline{\Gamma}$-$\overline{M}$ line form a bulk band inversion between Bi $p$ and Sm $d$ states. Intense surface bands forming Dirac cone dispersion are resolved at $\overline{\Gamma}$ and $\overline{M}$ high symmetry points. Ultra-high-resolution laser ARPES data from the high-quality single crystals enabled discovery of additional features in the electronic structure. The surface bands show opening of an energy gap and Fermi surface destruction across the Ne\'{e}l temperature in accord with our calculated results. The Dirac cone at $\overline{\Gamma}$ point is gapped at 15 K and the behavior remains similar across the magnetic transition. On the other hand, the Dirac cone at $\overline{M}$ appears to be gapless. The interplay between topology and magnetism is an interesting and open area of research that is still in its infancy. The results of SmBi presented here reveal an exotic scenario of momentum dependent gap formation at the Dirac point and destruction of one of the Fermi surfaces formed by topologically ordered surface states around $\overline{\Gamma}$ due to the long-range magnetic order which requires ideas beyond the existing theoretical models.

\section{Acknowledgements}

The authors thank the Proposal Assessing Committee, HiSOR (Prop. No.: 19BG053, 20AU007), N-BARD Hiroshima University for liquid Helium and Mr. Yogendra Kumar, School of Physical Sciences, Univ. Chinese Acad. Sciences, Beijing for help. Financial support from DAE, Govt. of India (Project No. RTI4003, DAE OM no. 1303/2/2019/R\&D-II/DAE/2079 dated 11.02.2020) is thankfully acknowledged. K.M. thanks DAE-BRNS for financial support under the DAE-SRC-OI Award program (21/08/2015-BRNS).

\end{document}